\documentclass[aps,pre,twocolumn,showpacs,groupaddress]{revtex4}
\usepackage{amsmath}
\usepackage{amsfonts}
\usepackage{amssymb}
\usepackage{epsfig}
\usepackage{graphicx}

\begin{document}

\title{Static point-to-set correlations in glass-forming liquids}

\author{Ludovic Berthier}
\affiliation{Laboratoire Charles Coulomb, UMR 5221, CNRS and Universit\'e
Montpellier 2, Montpellier, France}

\author{Walter Kob}
\affiliation{Laboratoire Charles Coulomb, UMR 5221, CNRS and Universit\'e
Montpellier 2, Montpellier, France}

\date{\today}

\begin{abstract}
We analyze static point-to-set  correlations in
glass-forming liquids. The generic idea is to freeze the position of a set
of particles in an equilibrium configuration and to perform
sampling in the presence of this additional constraint. Qualitatively different
geometries for the confining set of particles are considered
and a detailed comparison of resulting static and dynamic correlation functions
is performed. Our results reveal the existence of static 
spatial correlations not detected by conventional two-body
correlators, which appear to be decoupled from, and shorter-ranged than,  
dynamical length scales characterizing dynamic heterogeneity. We find
that the dynamics slows down dramatically under confinement, which suggests
new ways to investigate the glass transition. Our results indicate that
the geometry in which particles are randomly pinned is the best candidate 
to study static correlations.
\end{abstract}


\maketitle

The collective nature of the dynamics of
supercooled liquids approaching the glass transition is well 
established~\cite{reviewchi4}. A recent key advance was 
the study of multi-point dynamical
correlation functions instead of 
traditional two-point correlation functions such as the
intermediate scattering function. A central outcome is the determination
of a {\it dynamical} length scale increasing by a
factor of about 5-10 when glass-formers are cooled from normal liquid
conditions to temperatures around the glass temperature $T_g$.  
This raises the question of the underlying existence 
of nontrivial {\it structural} correlations,
which also grow when the glass transition is approached, and
how they relate to dynamic ones~\cite{rmp}. However, just as two-time
dynamic correlation functions do not detect directly dynamic
heterogeneity, two-point density correlation functions (pair correlation
functions) seem unable to capture
the relevant structural correlations, as these
functions only show a mild temperature dependence. There 
are proposals that three-body orientational 
order parameters might give insight for some
specific glass-formers~\cite{gilles,tanaka}, or that local
geometric structures might be significant~\cite{coslovich_07,jorge},
but {\it generic} methods to detect static
order still need to be devised.

Recently, the idea emerged that some form of 
`amorphous order' should develop in viscous liquids, which could 
be detected through `point-to-set' correlation 
functions~\cite{pointtoset,BB2,silvio,cavagna,rob}. Point-to-set (PTS) 
correlations probe static multi-point correlations,
since they are determined by fixing the position
of a `set' of $k$ particles and measuring the probability 
to find a $(k+1)$th particle at position ${\bf r}_{k+1}$.  
It can be hoped that if the geometry of the frozen set is well chosen, 
these multi-point functions yield spatial information 
without measuring how the correlation function 
depends on all its $k+1$ arguments. 
For a spherical cavity of radius $d$, for instance, 
one expects to detect a change of physical behavior when 
$d$ interferes with the relevant structural length scale~\cite{BB2}. 
Although first motivated in the context of the
random first order transition (RFOT) theory~\cite{rfot}, the set-up
is actually more general and does not rely on any hypothesis 
regarding the microscopic nature of the measured correlations. 
The generality of the approach thus strongly suggests 
that it is important to explore in detail these
PTS correlations in different geometries for the set as well as 
for various models of glassy systems.
Although reminiscent of studies of glass-formers in 
confined geometries~\cite{mckenna}, we
emphasize that PTS correlations probe genuine
{\it bulk} correlations, with no contribution from 
an external substrate~\cite{pointtoset,scheidler}. 
While difficult to implement for molecular systems, 
investigations along the lines suggested in the present
work could be performed in colloidal materials where it is possible
to freeze the position of a selected set of particles 
using for instance optical tweezers.

\begin{figure}[b]
\psfig{file=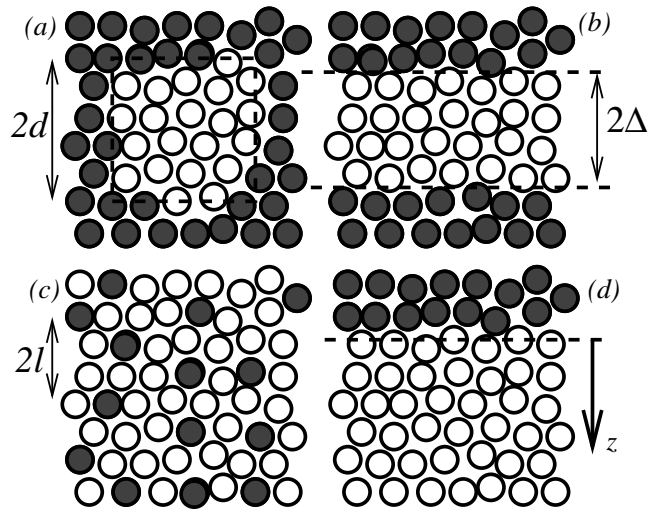,width=8.5cm,clip}
\caption{\label{example}
Four qualitatively 
different geometries to investigate point-to-set correlations
in glass-forming liquids.}
\end{figure} 

In this work, we show that PTS correlations can 
be detected using a broad variety of qualitatively distinct
geometries, see Fig.~\ref{example}, which
all reveal information on static correlations not included 
in conventional pair correlations. As a first step, 
we present the results of computer simulations of a simple 
glass-former in which we investigate how the geometry of 
the pinned particles influences the measured correlations
at a single state point. Our results suggest that the cavity 
geometry considered in earlier studies might not be the optimal choice
to study static correlations. Another surprising outcome of our analysis
is that static lengthscales appear to be decoupled from, and 
smaller than, dynamical lengthscales characterizing dynamic 
heterogeneity in the mildly supercooled regime typically 
studied in computer simulations.

For each of the geometries shown in Fig.~\ref{example}
we first equilibrate a three dimensional bulk system composed of $N$ particles.
At $t=0$ we permanently pin $k$ particles
(filled circles in Fig.~\ref{example}) whereas the remaining $N-k$
particles (open circles) move as before. Since the pinned particles
were chosen from the equilibrated fluid, the thermodynamics
of the free particles is strictly unperturbed, 
provided an average is performed over both thermal fluctuations and
different realizations of the pinning disorder~\cite{scheidler,krako}.
For the wall shown in Fig.~\ref{example}d, the
average density profile is for instance strictly 
constant, $\langle \rho(z) \rangle = \rho_0$, with no layering.
We use constant temperature molecular dynamics simulations to study 
a 50:50 binary mixture of harmonic spheres with size ratio 1.4, all
particles having the same mass $m$~\cite{tom}.  The unit of length
is given by $\sigma$, the diameter of the small particles, 
the unit of time
by $\sqrt{m \sigma^2/\varepsilon}$, where $\varepsilon$ is the
interaction strength, and temperature in units of $10^{-4} \varepsilon$,
setting the Boltzmann constant $k_B=1.0$.  We work at fixed density
$\rho_0=0.675$. For these parameters~\cite{sandalo}, 
slow dynamics sets in when $T
\lesssim 10$, a fit to a mode-coupling 
divergence yields $T_{c} \approx 5.2$, and the harmonic spheres
behave as quasi-hard spheres, as discussed in detail
in Refs.~\cite{tom}. Thus our system is a canonical model for studies
of the glass transition phenomenon. As announced, 
we present a comparative study of  the various confinement 
shown in Fig.~\ref{example} for a single, moderately low temperature, $T=8
> T_c$, a temperature at which the intermediate scattering function of
the bulk system already shows two-step relaxation.

In all cases we wish to answer the following question: How does the
presence of a pinned set of particles affect the structure and dynamics
of the remaining free particles? To quantify these effects we define two
overlap functions, akin to the collective and self intermediate scattering
functions. The collective overlap reads
\begin{equation}
Q(t) = \sum_i \left\langle n_i(t) n_i(0) \right\rangle / 
\sum_i \langle n_i(0) \rangle,
\label{q}
\end{equation}
where the sum runs over the cells (of volume $v\approx 0.53^3$,
comparable to the particle volume) of a cubic grid,
and $n_i(t) \in \{0,1\}$ is the occupation number of cell $i$ at time
$t$. We set $n_i=0$ if cell $i$ contains a pinned particle. The
overlap $Q(t)$ is close to 1 if configurations at time $0$ and $t$ 
are similar, but $Q(t)$ is unaffected by particle exchanges.
The long-time limit of the overlap,
$Q_\infty \equiv  Q(t\to\infty)$, provides 
direct information on static correlations. 
We also define the single particle function
\begin{equation}
Q_{\rm self}(t) = \sum_i \left\langle n_i^s(t) n_i^s(0) \right\rangle
/ \sum_i \langle n_i^s(0) \rangle,
\label{qself}
\end{equation}
where $n_i^s(t)=1$ if the same particle occupies the cell $i$
at times $0$ and $t$, and $n_i^s(t)=0$ otherwise.

We now describe the various geometries of
Fig.~\ref{example}. (a) Particles outside a cubic cavity of
linear size $2d$ are frozen.  The overlaps are measured at the center of
the cavity, using the $4^3$ central cells to improve the statistics;
8 independent realizations are studied for each $d$. A
similar (spherical) geometry has been studied in \cite{cavagna,rob}.
(b) Particles outside the range $0 < z < 2 \Delta$ are frozen, such
that the free particles are `sandwiched' between two infinite walls
separated by a distance $2\Delta$. The overlap is averaged over cells
located in the plane parallel to the walls in the middle of the 
sandwich; 10 independent realizations are studied for each $\Delta$,
with $L_x=L_y=13.68$.  (c) A finite fraction, $c$, of particles is
randomly selected in the fluid, with a typical distance between them 
$2 l = c^{-1/3}$~\cite{pinned}. The overlap is averaged over all cells. 
By using a large isotropic system, $L=16.3$, 
only few realizations (typically 2-3) are
needed to yield statistically accurate results.  
(d) Particles in the semi-infinite space $z<0$ are frozen. The overlap
is averaged over cells belonging to planes parallel to the wall at
distance $z$ from it.  This geometry has been studied in more detail 
(including different temperatures) in \cite{sandalo}.
For (b) and (d) we also included a hard wall at
the boundary between the wall(s) and the fluid to prevent free particles
to penetrate the walls~\cite{scheidler}. To perform quantitative 
comparisons, we define a `confining length' $\xi$, respectively as
$\xi = d$, $\Delta$, $l$, or $z$: a smaller $\xi$ means
stronger confinement. 
Physically, $\xi$ represents for each geometry the typical distance
between the point where the overlap is measured to the pinned set
of frozen particles.
Note finally that geometry (a) is peculiar 
since the number of confined particles is always finite, while it scales
with system size and diverges in the thermodynamic limit
in cases (b-d).

\begin{figure}[htb]
\psfig{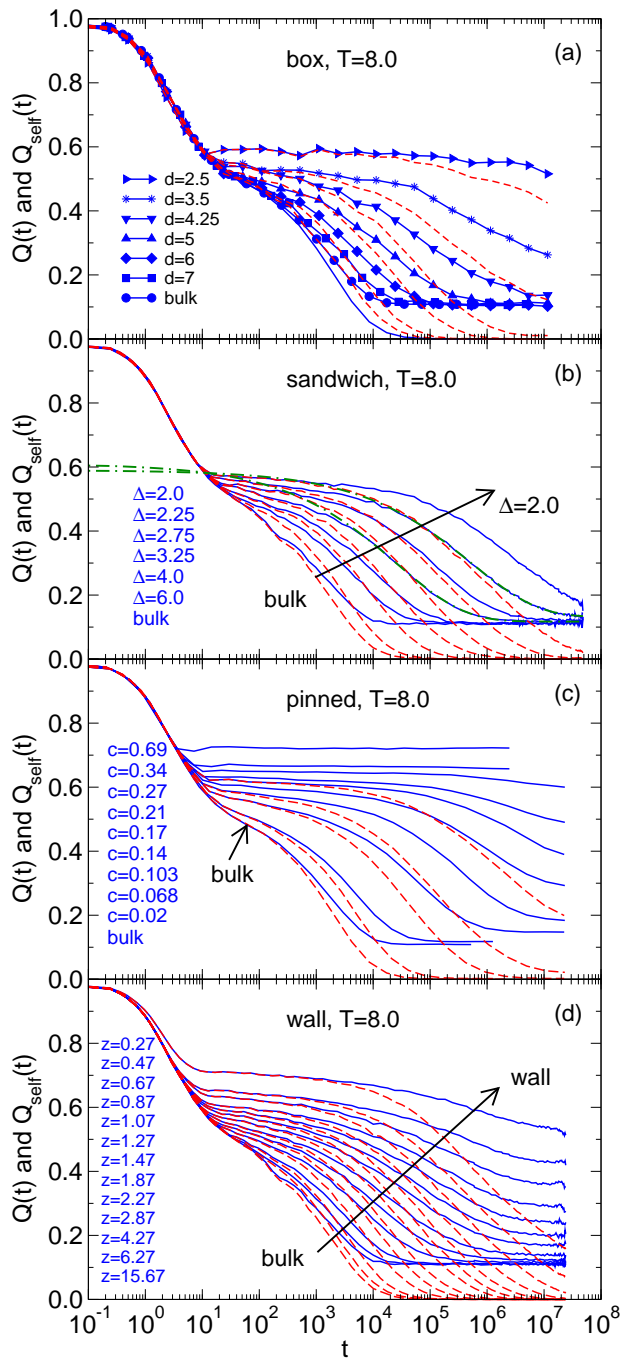}
\caption{\label{fig2}
Time dependence of the overlaps $Q(t)$ (full lines)
and $Q_{\rm self}(t)$ (dashed lines)
for $T=8.0$. The panels
correspond to the four geometries of 
Fig.~\ref{example}. In (b) we have also included the results of fits
with a stretched exponential to the $Q(t)$ data for $\Delta=2.25$
and $\Delta=3.25$ (dashed-dotted lines).}
\end{figure}

\begin{figure}[htb]
\psfig{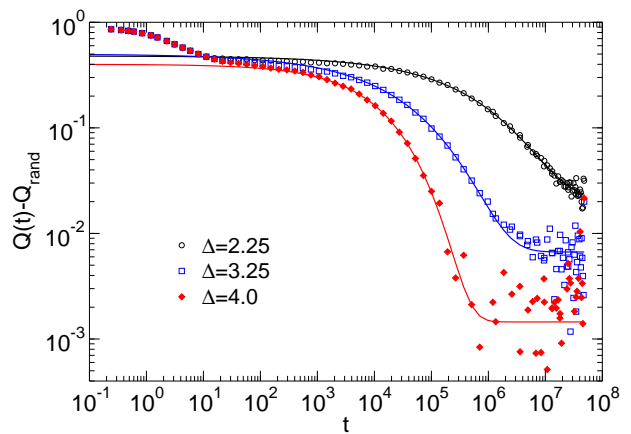}
\caption{\label{fig2_bis}
Time dependence of the overlap $Q(t)-Q_{\rm rand}$ (symbols)
and the stretched exponential fits (lines)
for the sandwich geometry and different values of $\Delta$.}
\end{figure}

In Fig.~\ref{fig2}  we gather our results 
for the four geometries, both overlaps (\ref{q})
and (\ref{qself}), and various degrees of confinement.
In all geometries the time correlation
functions have a similar qualitative behavior. When $\xi \to \infty$, 
bulk behaviour of the overlap $Q(t)$ is recovered, with
a two-step decay, and 
at long times a relaxation to the random value, 
$Q_{\rm rand} \approx 0.110595 \approx \rho_0 v$. When confinement
increases, the time dependence of $Q(t)$ slows down, while the long-time
limit increases, $Q_\infty(\xi) > Q_{\rm rand}$.
In practive we extract $Q_\infty(\xi)$ by fitting the long-time decay 
of $Q(t)$ to stretched exponential form. The quality of the fit 
is very good, as examplified in Fig.~\ref{fig2}b. In order to see that such
a fit does indeed allow to obtain $Q_\infty(\xi)$ with high precision, we show
in Fig.~\ref{fig2_bis} the time dependence of $Q(t) - Q_{\rm rand}$
on a logarithmic scale as well as the corresponding fits.
Finally we note that the evolution of $Q_{\rm self}(t)$ 
is similar, showing a dramatic
slowing down with increasing confinement, but its long-time limit is
always zero.

A comparison between both functions shows that $Q(t)$ roughly reaches 
$Q_\infty$ when $Q_{\rm self}(t)$ approaches zero, i.e.~essentially
when all particles have escaped the position they occupy at $t=0$. 
Thus, when $Q_\infty  >
Q_{\rm rand}$, non-random local density fluctuations persist
even though particles diffuse and
explore the available space. By monitoring the $\xi$-dependence of 
$Q_\infty$, we can
quantify the amount of static order imposed
by the confinement and have direct quantitative access to the influence of
the set of frozen particles on the fluid structure to obtain 
bulk, equilibrium, many-body information not contained in pair 
correlations~\cite{BB2,pointtoset,cavagna,silvio,rob,sandalo}.

>From Fig.~\ref{fig2}, we also conclude that measuring the evolution
of the static overlap $Q_\infty$ is more difficult than previously
thought~\cite{BB2,cavagna,cavagna_10}, because the dynamics slows down considerably 
with increasing confinement~\cite{scheidler}. Whereas the bulk
dynamics at $T=8$ corresponds to a moderately viscous state, there
exists in all four geometries a maximal confinement above which $Q(t)$
does not reach its long-time limit in the time window of our simulations,
and hence large values of $Q_\infty$ cannot be accessed. This 
difficulty gets even more pronounced at lower $T$~\cite{sandalo},
meaning that the measurement of point-to-set correlations closer
to $T_c$ is, at present, extremely challenging.

\begin{figure}
\psfig{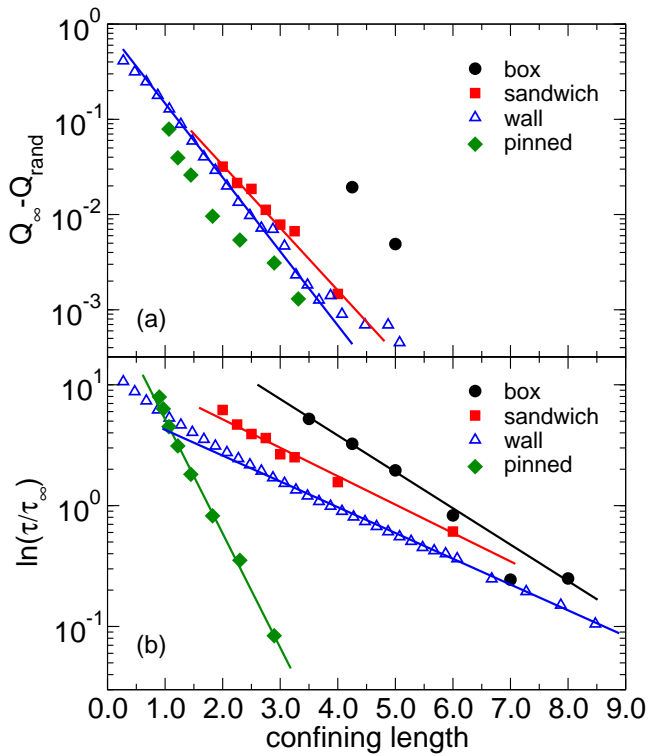}
\caption{\label{qinf4} 
a) Dependence of the overlap $Q_\infty$ on the confining length for all
geometries at $T=8$ (symbols) fitted with an exponential decay at long
distances (straight lines). b) Relaxation time of the self overlap
normalized by its bulk value.}
\end{figure}  

In Fig.~\ref{qinf4}, we present the
evolution of the average overlap $Q_\infty(\xi)$ as a function of the
confining length $\xi$ in all geometries (a), and the evolution
of the corresponding relaxation times $\tau(\xi)$ (b), defined
from the time decay of $Q_{\rm self}(t)$ to the value $1/e$. In the latter
figure we normalize the data by the bulk value 
$\tau_\infty = \tau(\xi \to \infty)$. The static
profiles confirm that the average overlap becomes increasingly non-random
by increasing the confinement since at a given confinement length we find that
$Q_{\infty}$ increases for the sequence pinned, wall, sandwich, box. Note that 
for the first three geometies the values of $Q_{\infty}$ are similar, whereas
the ones for the box are significanly larger. This indicates that 
the effect of confinement is highly non-linear.
In cases (b) and (d) we find that the large $\xi$ decay of $Q_\infty$ 
is compatible with an exponential decay, 
$Q_\infty(\xi) - Q_{\rm rand} \approx \exp ( -\xi / \xi_{\rm stat})$, 
which defines a PTS static lengthscale, $\xi_{\rm stat}$. 
For geometry $(d)$ this dependence holds for temperatures 
as low as $T=5.0 < T_c$~\cite{sandalo}.
A compressed exponential
decay was reported for a closed cavity in Ref.~\cite{cavagna},
which is unfortunately the geometry for which our data
is the most limited. For randomly pinned particles, 
one expects the non-random part of the overlap to 
scale as $c = 1/\xi^3$ for large $\xi$, and an exponential decay is not 
expected in that case, which requires a separate
analysis (see also \cite{gillespatrick}).

Regarding dynamics, we find that decreasing $\xi$
at constant $T$ leads to a strong slowing down of the dynamics
for both self and collective quantities.
In absolute values, $\tau(\xi)$ 
strongly increases with the number of confining walls 
which shows that relaxation is non-linearly
suppressed by adding more constraints.
For a given confining length, pinned particles have of course 
much less impact on the dynamics as a single particle
has obviously less effect than an entire wall. 
For the cavity (a), the slowing down is in fact so dramatic that 
the range of $\xi$ where $Q_\infty$ can be measured 
is very small: The overlap is too small when 
$\xi$ is large, but dynamics is much too slow when $\xi$ is small, 
which only leaves a narrow range to measure the static profile.
We conclude that measuring PTS correlations 
in a closed cavity is in practice a difficult task in the 
interesting supercooled regime.

For a single wall, $\tau(\xi)$ can be
followed down to $\xi \to 0$, while for the other geometries small values
of $\xi$ are not accessible due to a much stronger 
slowing down. All dynamic profiles
in Fig.~\ref{qinf4}b are well described 
by an exponential decay,
$\ln (\tau / \tau_\infty) \approx \exp(-\xi/\xi_{\rm dyn})$
which directly allows one to extract 
a dynamic correlation lengthscale $\xi_{\rm dyn}$~\cite{scheidler,tanaka2}.
In Ref.~\cite{sandalo} we have explored its temperature 
dependence for a single 
wall and discussed how $\xi_{\rm dyn}$
relates to previous measures of dynamic lengthscales.
>From Fig.~\ref{qinf4}b we see that the slowing down of the
dynamics is the least pronounced for the pinned geometry. Therefore we 
suggest that this geometry might be best suited best for the investigation
of the $T-$dependence of static correlations.

By comparing the two panels in Fig.~\ref{qinf4}, 
it is obvious that for all geometries the dynamics
seems to be affected over a broader range of confinement than
statics, which suggests that, generically, $\xi_{\rm dyn} > \xi_{\rm stat}$.
In fact we have frequently found parameters for which 
clear deviations from bulk dynamics are observed while the 
static overlap is still random, suggesting a clear decoupling
between static and dynamic correlations.
This decoupling is strong for a single wall and becomes 
less pronounced for randomly pinned particles. Our study thus seems to
confirm~\cite{silvio,sandalo}
that static correlation length scales are generically decoupled 
from dynamic ones and smaller, at least 
over the range of temperature we can explore numerically.

We finally discuss our results in a broader context. Theoretical progress
on the glass problem is slowed by the lack of
an obvious structural indicator to distinguish a glass
from a fluid. It has only recently been realized that, in the framework of
RFOT theory, the `ideal glass' phase below the Kauzmann
transition, $T_K$, is characterized by an infinite static PTS 
lengthscale~\cite{BB2,silvio}. This implies for instance that below $T_K$, 
freezing a semi-infinite space as in Fig.~\ref{example}d determines 
the position of the particles in the entire $z>0$ space~\cite{jorge}.
Spatially extended static profiles, such as shown in Fig.~\ref{qinf4}, 
thus uniquely characterize fluids approaching the glass 
transition~\cite{cavagna,sandalo}, and do not provide
valuable information for liquids at high temperatures.

Within RFOT theory, a static length $\xi_{\rm RFOT}(T)$
emerges from a spatial interpretation of the (mean-field) 
concept of metastable states~\cite{rfot}.
This `mosaic' length scale plays a role similar
to that of a nucleation length scale in first order transitions, because it
is set by the competition between the entropic gain of exploring different
states and the energy cost of having interfaces between them~\cite{rfot}.
This should be reflected, in the closed cavity of Fig.~\ref{example}a, as 
a crossover between a small $Q_\infty$ when $d > \xi_{\rm RFOT}$,
to a large $Q_\infty$ when $d < \xi_{\rm RFOT}$, since for decreasing
confinement the surface tension eventually dominates~\cite{BB2}.
A qualitatively similar crossover holds for
geometries (b, c) as well, but not for a single wall in (d)
where bulk behaviour with $Q_\infty = Q_{\rm rand}$ is recovered 
far from the wall and an interface must always be present.

In contrast to closed cavities,
the crossover set by $\xi_{\rm RFOT}$ involves in geometries (b) and
(c) a number of particles that diverges in the thermodynamic limit.
This `crossover' should therefore more properly be 
described as a genuine freezing transition towards 
an ideal glass phase where $Q_\infty$ is large
and density fluctuations do not relax (see also~\cite{chiara}). 
A glass phase can then be approached either
by decreasing $T$ in the bulk, or 
by increasing the confinement at constant $T$, which opens
exciting perspectives to study the glass transition. We are
presently pursuing the exploration of the glass phase 
obtained at large pinning density, but the results
go much beyond the theme of the present article and will
be presented elsewhere.
Our results also motivate further analysis 
of the phase diagram of viscous liquids
in confined geometries (b, c). Analytical calculations for 
randomly pinned particles exist for hard 
sphere systems within mode-coupling theory~\cite{krako-mct}, 
while the present results have motivated both
an RFOT analysis~\cite{chiara} and some 
numerical investigation~\cite{gillespatrick}.
Future work should characterize and compare in more detail 
the temperature evolution 
of the static and dynamic lengthscales introduced 
in this work beyond the case of the single wall studied in \cite{sandalo}. 

\acknowledgments

We thank G. Biroli, C. Cammarota,
D. Coslovich, R. Jack, J. Kurchan, D. Reichman, F. Sausset, G. Tarjus
and the authors of \cite{cavagna} for discussions.
W. K. is a member of the Institut universitaire de France.

\end{document}